\documentclass[preprint,preprintnumbers,amsmath,amssymb,12pt,floatfix,epsfig,nofootinbib,superscriptaddress]{revtex4}
\usepackage[usenames, dvipsnames]{color}
\usepackage{graphicx}
\usepackage{amssymb}
\usepackage[normalem]{ulem}
\renewcommand\sout{\bgroup\color[rgb]{1,0,0} \ULdepth=-.5ex \ULset}

\begin{document}
\title{Quasielastic Charged-Current Neutrino-Nucleus Scattering with Nonrelativistic Nuclear Energy Density Functionals}
\author{Kyungsik Kim}
\email{kyungsik@kau.ac.kr}
\affiliation{School of Liberal Arts and Science, Korea Aerospace University, Goyang 10540, Korea}
\author{Hana Gil}
\email{khn1219@gmail.com}
\affiliation{Center for Extreme Nuclear Matters, Korea University, Seoul 02841, Korea}
\author{Chang Ho Hyun}
\email{hch@daegu.ac.kr}
\affiliation{Department of Physics Education, Daegu University, Gyeongsan 38453, Korea}

\date{\today}
\begin{abstract}
Charged-current neutrino-nucleus scattering is studied in the quasielastic region
with the KIDS (Korea-IBS-Daegu-SKKU) nuclear energy density functional.
We focus on the uncertainties stemming from the axial mass and the in-medium effective mass of the nucleon.
Comparing the result of theory to the state-of-the-art data from MiniBooNE, T2K, and MINER$\nu$A, 
we constrain the axial mass and the effective mass that are compatible with the data.
We find that the total cross section is insensitive to the effective mass, 
so the axial mass could be determined independently of the uncertainty in the effective mass. 
Differential cross sections at different kinematics are, on the other hand, sensitive to the effective mass as well as the axial mass.
Within the uncertainty of the axial mass constrained from the total cross section, dependence on the effective mass is examined.
As a result we obtain the axial mass and the effective mass that are consistent with the experimental data.
\end{abstract}

\maketitle

\section{Introduction}

Measurement of the neutrino-nucleus ($\nu-A$) scattering cross section in the last decade at MiniBooNE \cite{mini1, mini2, mini3, mini4,kdar},
MINER$\nu$A \cite{mine1,mine2, mine3, mine4} and T2K \cite{t2k1, t2k2} has improved the accuracy of the data dramatically, so the era of precision neutrino physics is dawning.
One major purpose of the experiments is to resolve long-standing puzzles such as the neutrino mass, flavor oscillation, and CP violation in the leptonic sector.
Success of the forthcoming experiments is expected to identify the limit of the standard model more stringently and lead to a new physics beyond the standard model.
Interaction of the neutrino with nuclei plays a crucial role in understanding the result of the experiment.
For a precise measurement of the standard model physics, uncertainties stemming from both hadronic and nuclear structures should be understood correctly, and should be reduced as much as possible.
Those uncertainties also play a critical role in the interaction of neutrinos with nuclear matter at finite density and temperature, which has an essential consequence in the explosion of supernovae, and thermal evolution of the neutron star.

KIDS (Korea-IBS-Daegu-SKKU) nuclear energy density functional (EDF) was initiated with a prospect to construct a nuclear model in which finite nuclei and infinite nuclear matter can be described to desired accuracy within a single framework.
A series of works applied the model to nuclear matter and nuclei \cite{kids-nm, kids-nuclei1, kids-nuclei2}.
The results showed that a unified description of nuclei and nuclear matter is feasible by expanding nuclear EDF in the power of the Fermi momentum.
Combining the nuclear data and the neutron star observations,
parameters in the symmetry energy could be constrained within narrow ranges \cite{kids-se, kids-k0, kids-ksym}.
Extending the range of application, we considered quasielastic electron scattering off nuclei with the nuclear wave functions
obtained with KIDS EDF \cite{kids-eAexcl, kids-eAincl}.
Without any adjustment of the model parameters to scattering data, KIDS EDF reproduces the experimental data accurately.
Uncertainties in the nuclear structure arising from the nucleon effective mass in nuclear medium and the symmetry energy have been explored in detail.
Some results turn out to depend on the effective mass sensitively,
so it is demonstrated that the electron scattering could be a tool to constrain the effective mass of the nucleon in nuclear medium.

Stimulated by the success in the electron scattering,
we apply the KIDS EDF to the $\nu-A$ scattering, and explore the uncertainty due to the in-medium effective mass of the nucleon and the axial mass, simultaneously.
Nuclear wave functions are obtained by solving Hartree-Fock equations in which nonrelativistic nuclear potentials are imported from the KIDS EDF.
Role of the effective mass is examined by using four models KIDS0, KIDS0-m*77, KIDS0-m*99 and SLy4, in which isoscalar and isovector effective masses at the saturation density are
$(\mu_s,\ \mu_v) =$ (1.0, 0.8), (0.7, 0.7), (0.9, 0.9) and (0.7, 0.8), respectively in the unit of free nucleon mass.
Axial mass is defined in terms of the form factor slope at four-momentum transfer $Q^2=0$ as $M_A = [G'_A(0)/2G_A(0)]^{-1/2}$ where $G_A(Q^2)$ denotes the axial form factor of the nucleon.
Dependence on the axial mass is considered by employing a standard value $M_A = 1.032$ GeV,
and a large value $M_A = 1.30$ GeV.

In the result we find that in several kinematic conditions, the effect of the effective mass appears to be clear, 
and the result agrees with data better when the isoscalar effective mass at the saturation density is close to the free mass.
On the other hand, when the difference due to the effective mass is small, theoretical results agree 
well with the data regardless of the effective mass.
Contribution of the axial mass is discriminated well in the total cross section of the neutrino.
Large axial mass $M_A=1.30$ GeV reproduces the MiniBooNE data better than the standard value $M_A=1.032$ GeV.
Interestingly the total cross section is insensitive to the effective mass,
so the role of the axial mass can be singled out and probed without being interfered by other uncertainties.
In the comparison of the differential cross section, large axial mass combined with large effective mass gives better agreement to data on average.
However more accurate measurements are demanded to constrain the axial mass with the differential cross sections.

In the present paper, the formalism of the charged-current (CC) $\nu-A$ scattering is briefly introduced in Sec. II,
and Section III presents the results and discussion.
Finally, we summarize the work in Sec. IV.

\section{Formalism}

The $\nu({\bar \nu})-A$ scattering is described by the connection of the electromagnetic interaction and weak interaction.
In order to calculate the $\nu({\bar \nu})-A$ scattering, we choose that the target nucleus is seated at the origin of the coordinate system.
$p_i^{\mu}=(E_i, {\bf p}_i)$, $p_f^{\mu}=(E_f, {\bf p}_f)$, $p_A^{\mu}=(E_A, {\bf p}_A)$, $p_{A-1}^{\mu}=(E_{A-1}, {\bf p}_{A-1})$, and $p^{\mu}=(E_N, {\bf p})$ represent the four-momenta of the incident neutrino, outgoing neutrino, target nucleus, the residual nucleus, and the knocked-out nucleon, respectively.
For the CC reaction in the laboratory frame, the inclusive cross section is given by the contraction between lepton and hadron tensors:
\begin{eqnarray}
{\frac {d\sigma} {dT_N}} &=& 4\pi^2{\frac {M_N M_{A-1}} {(2\pi)^3
M_A}} \int \sin \theta_l d\theta_l \int \sin \theta_N d\theta_N p
f^{-1}_{rec} \sigma^{W^{\pm}}_M [ v_L R_L  + v_T R_T + h v_T' R_T' ],
\label{cs}
\end{eqnarray}
where $M_N$ is the nucleon mass in free space, $\theta_l$ denotes the scattering angle of the lepton, $\theta_N$ is the polar angle of knocked-out nucleons, $T_N$ is the kinetic energy of the knocked-out nucleon, and $h=-1$ $(h=+1)$ corresponds to the intrinsic helicity of the incident neutrino (antineutrino).
The $R_L, R_T$ and $R^{'}_T$ are longitudinal, transverse, and transverse interference response functions, respectively.
Detailed forms for the kinematical coefficients $v$ and the corresponding response functions $R$ are given in Refs. \cite{kimprc08,kimprc16}.
The squared four-momentum transfer is given by $Q^2=q^2 - \omega^2=-q^2_{\mu}$.
For the CC reaction, the kinematic factor $\sigma^{W^\pm}_M$ is defined by
\begin{equation}
\sigma^{W^\pm}_M = \sqrt{1 - {\frac {M^2_l} {E_f}}} \left ( {\frac
{G_F \cos (\theta_C) E_f M_W^2} {2\pi (Q^2 + M^2_W)}} \right )^2,
\end{equation}
where $M_W$ is the rest mass of $W$-boson, and $M_l$ is the mass of an outgoing lepton.
$\theta_C$ represents the Cabibbo angle given by $\cos^2 {\theta_C} \simeq 0.9749$.
$G_F$ denotes the Fermi constant.
The recoil factor $f_{rec}$ is written as
\begin{equation}
f_{rec} = {\frac {E_{A-1}} {M_A}} \left | 1 + {\frac {E_N}
{E_{A-1}}} \left [ 1 - {\frac {{\bf q} \cdot {\bf p}} {p^2}}
\right ] \right |. \end{equation}


The nucleon current $J^{\mu}$ represents the Fourier transform of the
nucleon current density written as
\begin{equation}
J^{\mu}=\int {\bar \psi}_p {\hat {\bf J}}^{\mu} \psi_b e^{i{\bf
q}{\cdot}{\bf r}}d^3r,
\end{equation}
where ${\hat {\bf J}}^{\mu}$ is a free weak nucleon current operator,
and $\psi_{p}$ and $\psi_{b}$ are wave functions of the knocked-out and the bound state nucleons, respectively.
The wave functions are generated with the same approach as the previous work \cite{kids-eAincl}.
For a free nucleon, the current operator of the CC reaction consists of the weak vector and the axial vector form factors:
\begin{equation}
{\hat {\bf J}}^{\mu}=F_{1}^V (Q^2){\gamma}^{\mu}+ F_{2}^V
(Q^2){\frac {i} {2M_N}}{\sigma}^{\mu\nu}q_{\nu} + G_A(Q^2)
\gamma^{\mu} \gamma^5+ {\frac {1} {2M_N}}G_P(Q^2) q^{\mu}
\gamma^5. \label{relJ}
\end{equation}
By the conservation of the vector current (CVC) hypothesis, the vector form
factors for the proton (neutron), $F_{i}^{V,~p(n)} (Q^2)$, are expressed as
\begin{eqnarray}
F_i^{V} (Q^2) &=& F_i^{p} (Q^2) - F_i^{n}(Q^2). \label{weak-ff}
\end{eqnarray}

The axial form factors for the CC reaction are given by
\begin{eqnarray}
G_A (Q^2) &=& -g_A /(1+Q^2/M_A^2)^2,
\label{gs}
\end{eqnarray}
with $g_A=1.262$ and two values 1.032~GeV and 1.30~GeV are assumed for $M_A$.

The induced pseudoscalar form factor is parameterized by the
Goldberger-Treimann relation
\begin{equation}
G_P(Q^2) = {\frac {2M_N} {Q^2+m^2_{\pi}}} G_A(Q^2),
\end{equation}
where $m_{\pi}$ is the pion mass.
But the contribution of the pseudoscalar form factor vanishes for the neutral-current reaction because of
the negligible final lepton mass participating in this reaction.

\section{Result}
With the KIDS EDF model, we calculate the various differential cross sections and total cross sections in the quasielastic CC $\nu - A$ scattering off $^{12}$C, and compare the result with MiniBooNE, MINER$\nu$A, and T2K data.
In order to obtain the wave functions of bound and final nucleons from nonrelativistic nuclear model, the relativistic wave functions are generated by using the nonunitary transformation \cite{kids-eAexcl,kids-eAincl,clark,kelly}.
For the Coulomb distortion of the final lepton, the same approximation exploited by the Ohio group \cite{kimcoulomb} is used.
In these neutrino experiments, the energy of the incident neutrino cannot be fixed but has an energy spectrum, 
so the cross sections have to be averaged over the flux of the incoming neutrino beam.

\begin{figure}
\centering
\includegraphics[width=13cm]{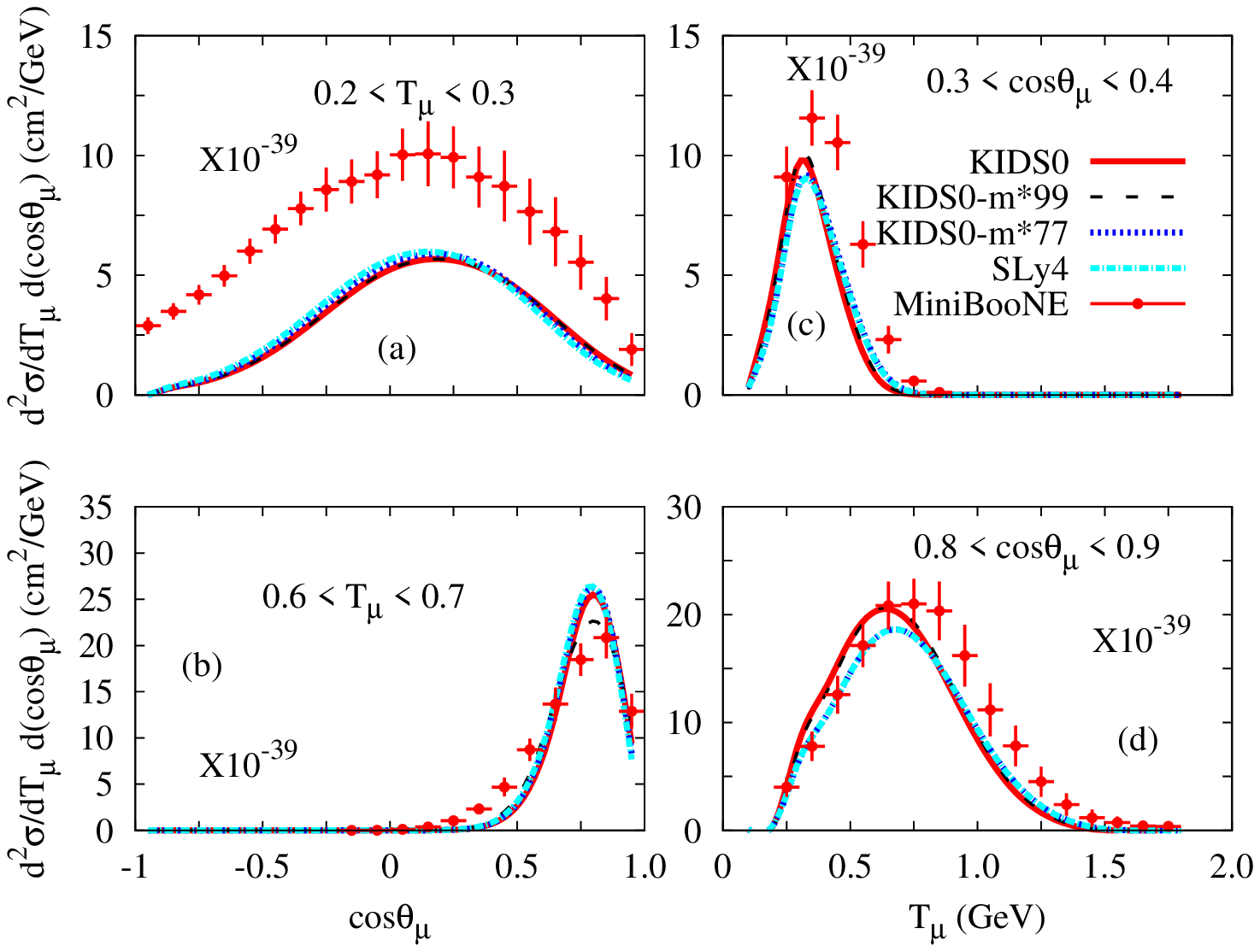}
\includegraphics[width=13cm]{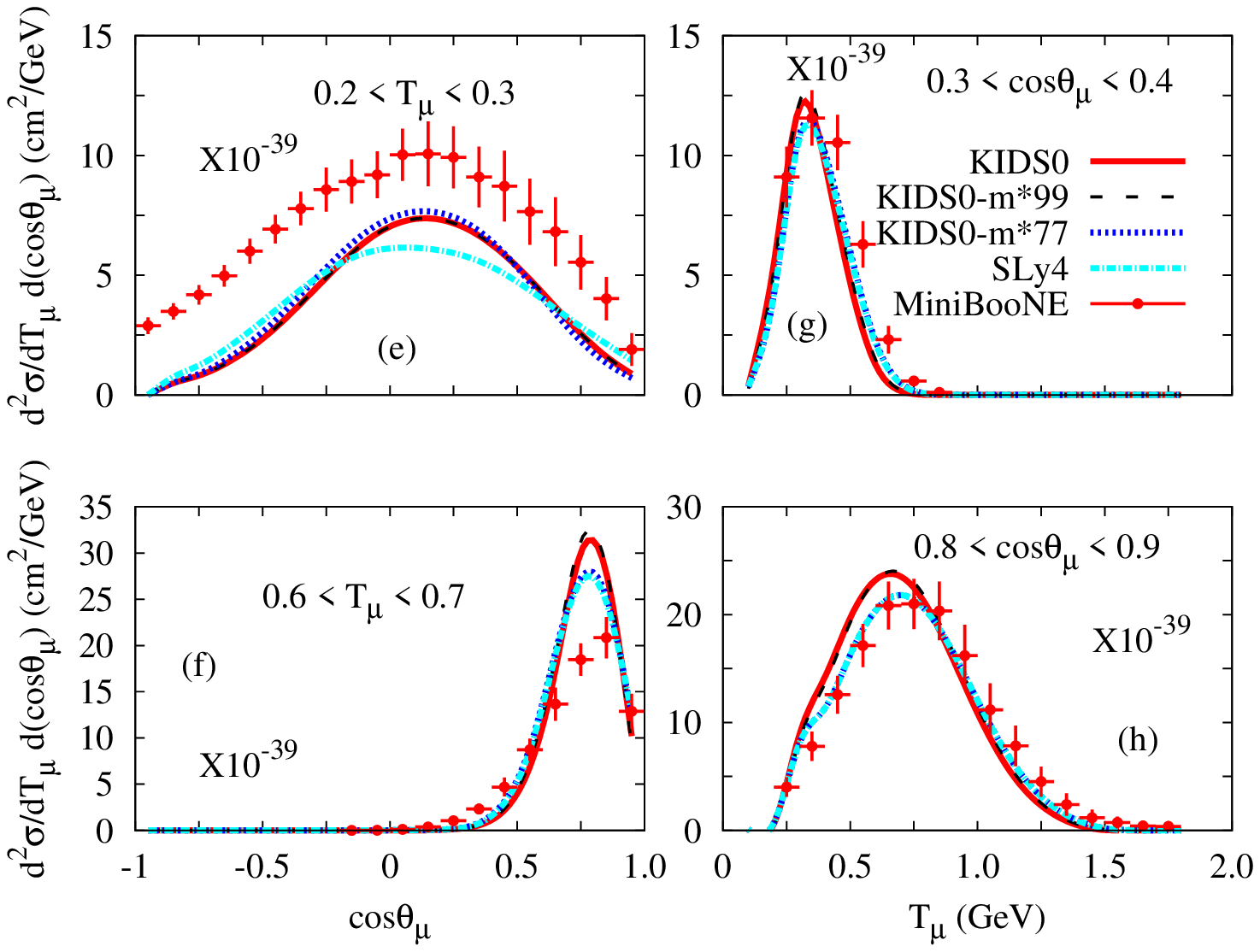}
\caption{Flux-averaged double-differential cross sections in terms of the kinetic energy and the scattering angle of the final muon from $^{12}$C.
The data were measured by the MiniBooNE Collaboration \cite{mini1}.
Panels (a-d) are the result with $M_A=1.032$ GeV, and (e-h) with $M_A=1.30$ GeV.}
\label{fig1}
\end{figure}

Figure \ref{fig1} shows the flux-averaged double-differential cross sections in terms of polar angle or kinetic energy of the outgoing muon.
Data in the given kinetimatic regions are available from the MiniBooNE Collaboration at Fermi Lab. \cite{mini1}.
The theoretical cross sections in Fig. \ref{fig1} (a) $\sim$ (d) and (e) $\sim$ (h) are the results for the value of the axial mass
$M_A=1.032$ and 1.30 GeV, respectively.
The theoretical results do not describe the data well at fixed $T_{\mu}$ (espeically $0.2 < T_\mu < 0.3$ GeV region),
but describe the data relatively well at fixed the angle.
The results with large axial mass provide better agreement but the influence of the effective mass cannot be distinguished in Fig. \ref{fig1}.

\begin{figure}
\begin{center}
\includegraphics[width=16cm]{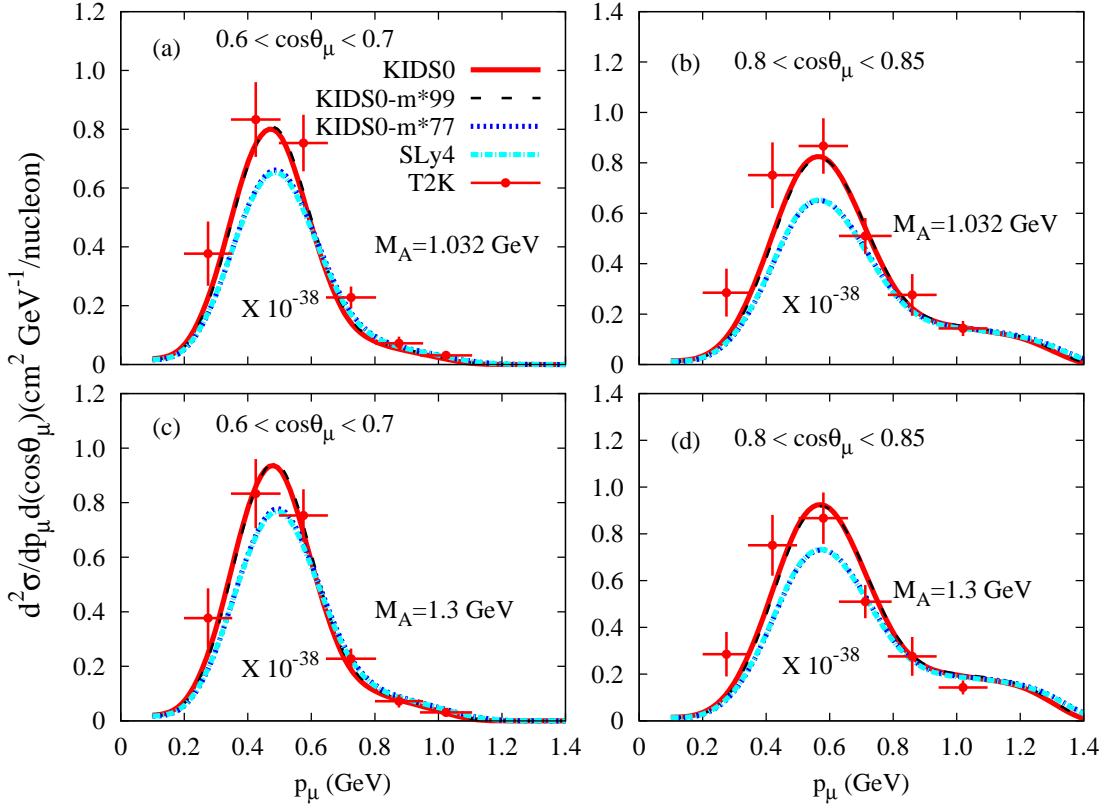}
\end{center}
\caption{Flux-averaged double-differential cross sections in terms of the incident muon momentum at fixed angle of muon.
The data were measured by the T2K Collaboration \cite{t2k1}.
Panels (a, b) are the result with $M_A=1.032$ GeV, and (c, d) with $M_A=1.30$ GeV.}
\label{fig2}
\end{figure}

In Fig. \ref{fig2}, the flux-averaged double-differential cross sections with kinematics different from Fig. \ref{fig1} are shown in terms of the momentum of the
outgoing muon at fixed polar angle of the muon and compared with the data measured from T2K \cite{t2k1}.
The value of $M_A$ is 1.032 GeV in the upper panels and 1.30 GeV in lower panels.
The results of $M_A=1.30$ GeV agree with the data better than the results of 1.032 GeV.
From these results, the effect of the effective mass appears to be clear,
and the results of the large effective mass agree with the data better than the small ones.

\begin{figure}
\centering
\includegraphics[width=15cm]{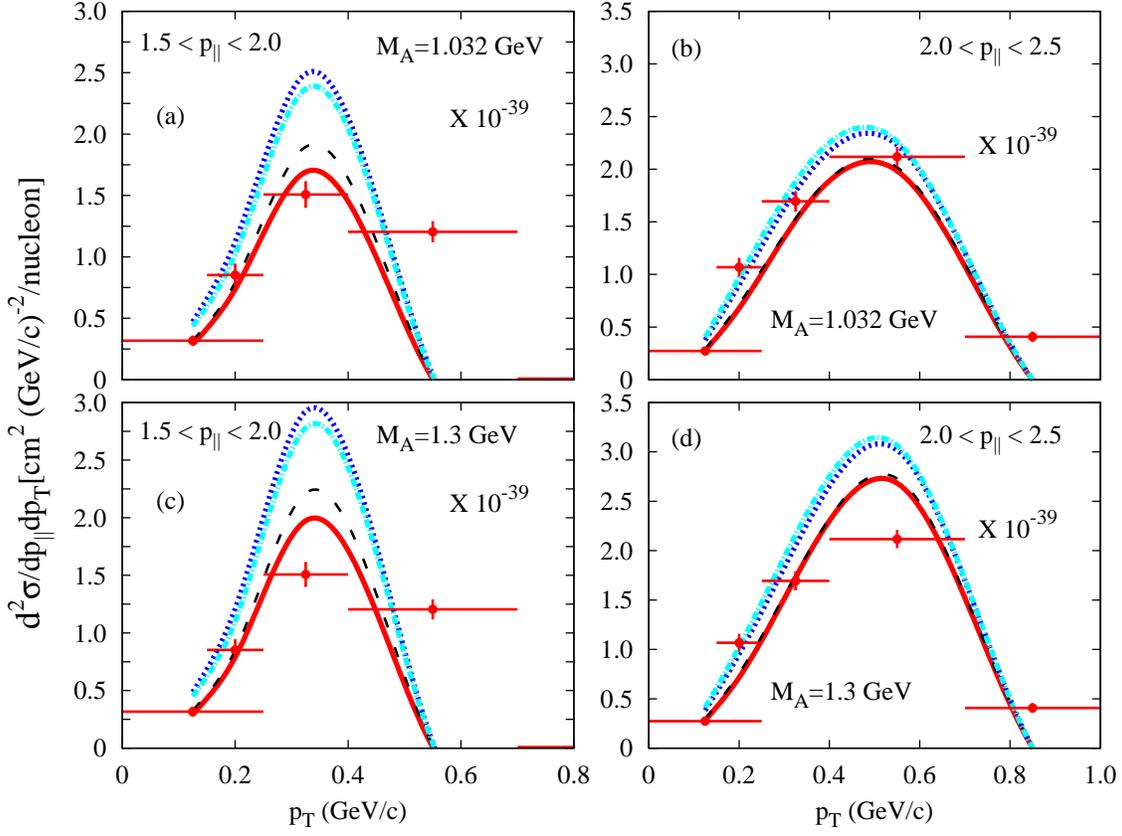}
\caption{Flux-averaged double-differential cross sections in terms of the muon transverse momentum at fixed longitudinal momentum.
The data were measured by the MINER$\nu$A Collaboration \cite{mine4}.}
\label{fig3}
\end{figure}

In Fig. \ref{fig3}, the flux-averaged double-differential cross sections are shown for the outgoing muon antineutrino in terms of $p_T$, where $p_T$ and $p_{\parallel}$ represent the transverse and longitudinal component of the muon momentum with respect to the incident antineutrino beam, respectively.
According to Ref. \cite{mine4}, 
this kinematics was exploited to include the nuclear effects in the $\nu - A$ scattering like the final state interaction, meson production, and so on.
In this work, we show the results of low momenta because the inelastic processes like meson production are excluded 
and the numerical difficulty is avoided due to partial-wave expansion.
The legend of the curve is the same as the Figs. \ref{fig1} and \ref{fig2}.
In this case, the theoretical results of $M_A=1.032$ GeV describe the data better than the results of $M_A=1.30$ GeV.
Contrary to the results in Figs. 1, 2, the large effective mass results are suppressed to the small effective mass ones in these kinematics, but agree with the measurement better than the small effective masses.

\begin{figure}
\centering
\includegraphics[width=15cm]{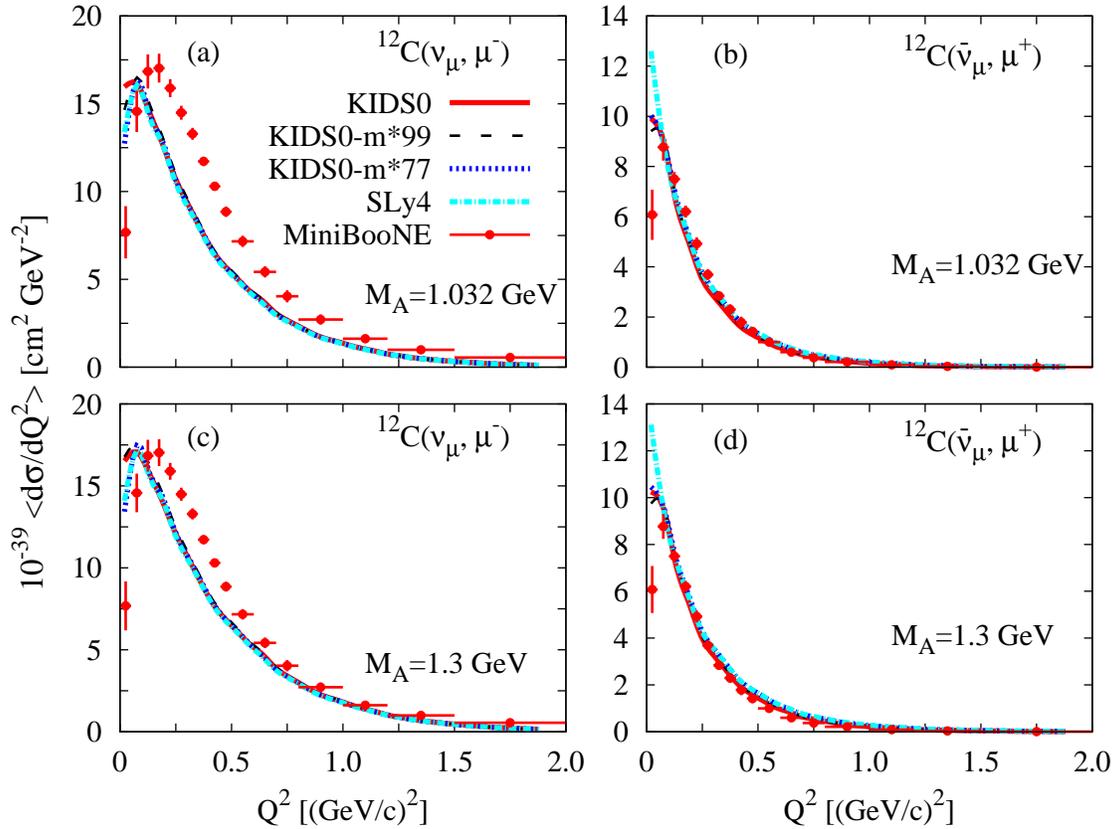}
\caption{Flux-averaged differential cross sections in terms of the four-momentum transfer squared.
The data were measured by the MiniBooNE Collaboration \cite{mini1}.}
\label{fig4}
\end{figure}

Figure \ref{fig4} shows the flux-averaged differential cross sections in terms of the squared four-momentum transfer 
for the incident neutrino and antineutrino scattering.
In the case of the incoming neutrinos, for both values of $M_A$, the theoretical results shift to the left side of data by about 0.1 (GeV/$c$)$^2$.
Axial mass tends to increase the magnitude of the cross section.
As a result, the result of $M_A=1.30$~GeV agrees to data better than $M_A=1.032$~GeV.
For the antineutrino, the data are reproduced well by both $M_A=1.032$ and 1.30~GeV.
On the other hand, the effective mass hardly affects the cross section in this kinematics.

\begin{figure}[t]
\centering
\includegraphics[width=15cm]{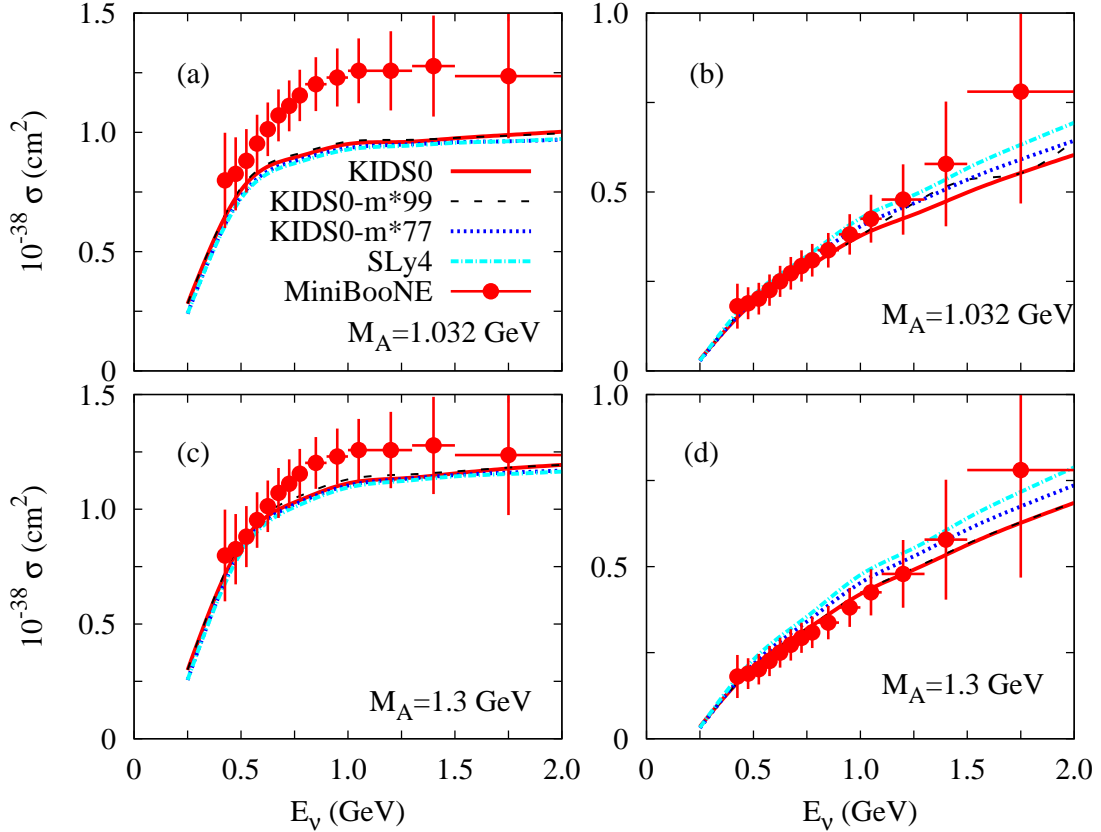}
\caption{Total scaled cross sections in terms of the incident neutrino energy.
Left panels are the result for the neutrino and the right panels for the antineutrino.
Upper panels are the result for $M_A=1.032$ MeV, and lower panels are the result for $M_A=1.30$ MeV.
The experimental data for the incident neutrino and antineutrino were measured from MiniBooNE \cite{mini1} and \cite{mini3}, respectively.}
\label{fig5}
\end{figure}

We calculate the total scaled cross sections in terms of the incident neutrino (antineutrino) energies, 
in which the total cross section is divided by the number of participated nucleon in the reaction.
The results are shown in Fig.~\ref{fig5}.
For the neutrino, the effect of the $M_A$ enhances the cross section about 30 \% but for the antineutrino it does about 15 \%.
In the case of the antineutrino, the influence of the effective mass increases the cross section at higher incident energies
although its effect is very small for the neutrino.

\begin{figure}[t]
\centering
\includegraphics[width=15cm]{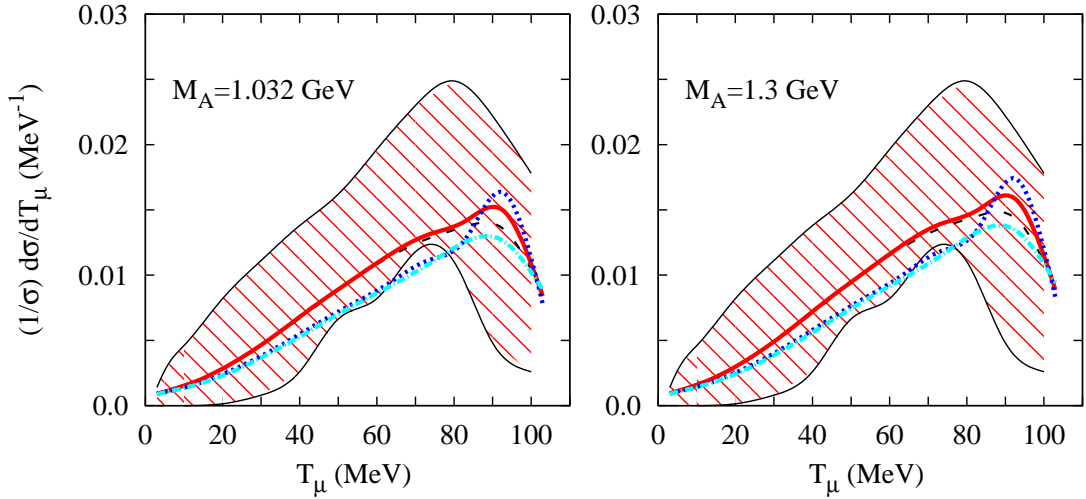}
\caption{Differential cross sections in terms of the kinetic energy of the outgoing muon.
Left panel is the result for $M_A=1.032$ GeV, and right panel for $M_A=1.30$ GeV.
The red solidus parts are obtained from KDAR data on MiniBooNE \cite{kdar}.}
\label{fig6}
\end{figure}

Recently, a new experiment was performed at MiniBooNE \cite{kdar} with monoenergetic muon neutrinos at 236 MeV, which are created when a positive kaon at rest decays, called kaon-decays-at-rest (KDAR).
We calculate the differential cross sections in terms of the kinetic energy of the outgoing muon and compare the result with the data in Fig.~\ref{fig6}.
The red solidus part is with shape-only $1\sigma$ error band, where $\sigma$ denotes total cross section and
yields $\sigma = (2.7 \pm 1.2) \times 10^{-39}$ cm$^2$/neutron.
The legend of the curve is the same as Fig. \ref{fig1}.
In the low incident neutrino energies, the effect of the $M_A$ is small, so gives a difference less than 10 \%.
The cross sections of small effective mass are out of the data around the peak for both $M_A=$1.032~GeV and 1.30~GeV.

\section{Summary}

Charged-current quasielastic scattering of the neutrino and antineutrino with $^{12}$C target has been considered within a nonrelativistic
nuclear density functional theory.
Both the wave functions of bound nucleon in $^{12}$C nucleus and final state interactions of the outgoing nucleons are obtained
by using the effective nuclear potentials obtained from KIDS EDF.
Parameters of the KIDS EDF have been fixed to satisfy well-defined nuclear matter properties and nuclear data,
and there is no calibration of the model to scattering data.

We found that the model reproduces experimental data at various kinematics very well.
At the same time, dependence on the in-medium effective mass of the nucleon and its axial mass is identified clearly.
Dependence on the effective mass is probed by using two groups of models,
one group with isoscalar effective mass close to the free mass ($\mu_s \simeq 1$),
and the other group with $\mu_s \simeq 0.7$.
Comparisons with the data from T2K and MINER$\nu$A Collaborations are crucial in diagnosing the effect of the effective mass.
Results of the KIDS EDF are in good agreement with the T2K and MINER$\nu$A data with $\mu_s\simeq 1$.
It is also confirmed that the dependence on the effective mass is dominated by the isoscalar effective mass,
and the role of the isovector effective mass can be neglected.
We observed the same behavior in the quasielastic electron scattering, in which $\mu_s\simeq 1$ models agree with the data better than the $\mu_s \simeq 0.7$ models.

In the comparison with the MiniBooNE data,
role of the effective mass becomes less dominant compared to the T2K and MINER$\nu$A data,
but the effect of the axial mass becomes crucial.
A highlighting result is the total cross section of the neutrino,
where the standard value of the axial mass $M_A = 1.032$ GeV fails to reproduce the neutrino data.
With $M_A=1.30$ GeV, theory results reside within the experimental uncertainty.
It is notable that the total cross section of the antineutrino is insensitive to $M_A$.
More importantly, total cross sections depends on the effective mass very weakly,
so they provide a unique opportunity to constrain the uncertainty of the axial mass.
The effect of the axial mass $M_A$ is small at low incident energies of the neutrino,
but the effect increases with higher neutrino energies.
It is argued that the axial mass could be interpreted to play a role to subsume the higher-order contributions
such as the multi-meson-exchange currents or multi-particle-multi-hole processes.
We assumed impulse approximation in the calculation.
It seems that large $M_A$ values are favorable if the transition matrix elements are evaluated in the impulse approximation.

\section*{Acknowledgments}
This work was supported by the National Research Foundation of Korea (NRF) grant funded by the Korea government (No. 2018R1A5A1025563 and No. 2020R1F1A1052495).

\end{document}